\documentstyle[prl,aps,twocolumn,psfig,floats]{revtex}
\input epsf
\draft
\begin{document}
\twocolumn[\hsize\textwidth\columnwidth\hsize\csname@twocolumnfalse%
\endcsname

\title{Orbital liquid in three dimensional Mott insulator: $LaTiO_3$}
\author{G.Khaliullin${^{1,2}}$, and S.Maekawa ${^2}$}
\address {${^1}$ Max-Planck-Institut f\"ur Festk\"orperforschung,
  Heisenbergstr.1, D-70569 Stuttgart, Germany \\
  ${^2}$Institute for Materials Research, Tohoku University, Sendai 980-8577, Japan}
\maketitle
\begin{abstract}
We present a theory of spin and orbital states in Mott insulator
$LaTiO_3$. The spin-orbital superexchange interaction between $d^1(t_{2g})$
ions in cubic crystal suffers from a pathological degeneracy 
of orbital states at classical level. Quantum effects remove this 
degeneracy and result in the formation of the coherent ground state, 
in which the orbital moment of  $t_{2g}$ level is fully quenched.
We find a finite gap for orbital excitations.
Such a disordered state of local degrees of freedom on unfrustrated, 
simple cubic lattice is highly unusual. Orbital liquid state   
naturally explains observed anomalies of $LaTiO_3$.

\end{abstract}
\pacs{PACS numbers: 75.10.-b, 71.27.+a, 75.30.Et, 75.30.Ds}
]

\def\bottomfraction{.9}
\def\textfraction{.1}

Experiments on transition metal oxides continue to challenge the theory -
Keimer et al. \cite{kei00} 
have recently reported on dynamical quenching of
$t_{2g}$ orbital angular moments in Mott insulator $LaTiO_3$ which is not associated by any detectable orbital/Jahn-Teller ordering.
More specifically, magnon spectra are found to fit a spin one-half, nearly
isotropic Heisenberg model on cubic lattice. At the same time, anomalous x-ray 
scattering study has not revealed any evidence of static orbital order.
These observations indicate that  $t_{2g}$ orbital degrees of freedom
of $Ti^{3+}$ ions are mysteriously missing in a low energy states of
$LaTiO_3$. Keimer et al. suggested a picture of strongly fluctuating 
orbitals to reconcile their results.

Dynamical quenching of local orbital moments in a periodic, simple cubic 
lattice formed by  $Ti^{3+}$ ions in $LaTiO_3$ is fascinating. 
This poses a serious problem for the canonical Goodenough-Kanamori picture
of successive orbital and magnetic orderings - the guiding idea in the 
modern theory of orbitally degenerate transition metal oxides 
(see for review \cite{tok00,kug82}). Loosely speaking, a lifting of the 
orbital degeneracy without symmetry breaking is formally similar to the  
spin moment quenching in periodic Kondo lattices. The novelty of 
$LaTiO_3$ is however that it is an insulator. At the same time, 
the ideas of quantum disorder due to the weak connectivity 
or geometrical frustration (like in pyrochlore systems, see \cite{can98} for 
instance) do not apply here, either. 

As discussed below, there are several physical reasons for 
the orbital disorder in $LaTiO_3$, and the 
key factor is a special symmetry of $t_{2g}$ -superexchange interaction 
on cubic lattice. This results in a quantum resonance between orbital 
levels, which removes degeneracy present at classical level by the formation
of resonating bonds in orbital sector - a realization of Anderson's 
resonating valence bonds (RVB)
idea \cite{and87} in a three dimensional (3D) insulator with help of 
orbital degrees of freedom. As for spin moments, they show a 
long-range order as expected on general grounds. However, a staggered 
moment is considerably reduced, since short-wavelength magnons are 
actively involved in the local resonance of exchange bonds - in fact,
it is the composite spin-orbital excitation which plays a crucial 
role in the theory presented.

We begin with discussion of the $t_{2g}$ -superexchange interaction.
In terms of fermions
$a_{i,\sigma}$, $b_{i,\sigma}$, $c_{i,\sigma}$
corresponding to $t_{2g}$ levels of
$yz$, $zx$, $xy$
symmetry, respectively, the hopping term in the Hubbard model 
reads as
$-t\left(a^\dagger_{i,\sigma}a_{j,\sigma} +
b^\dagger_{i,\sigma}b_{j,\sigma} + {\rm H. c.}\right)$
for the bonds along {\it c}-direction in a cubic crystal. Correlation
energies in doubly occupied virtual states of $Ti^{3+}$ ion 
are specified as follows: $U$ for electrons on the same orbital; 
$U^{'}+J_H$, if they occupy different orbitals and form a
spin-singlet; and $U^{'}-J_H$ for a spin-triplet charge excitation. 
The relation $U^{'}=U-2J_H$ holds in the atomic limit. 
As  $J_H << U$ usually,
the dominant part of the superexchange interaction in a cubic 
lattice is:
\begin{equation}
H_{\rm SE}= -\frac{4t^2}{U} + \sum_{\left\langle ij \right\rangle}
\left({\bf S}_i{\bf S}_j +\frac{1}{4}\right)\hat J_{ij}^{(\gamma)} ~.
\end{equation}
On {\it c}-axis bonds the orbital structure of the exchange ``constant''
is given by (hereafter, the energy scale $4t^2/U $ is used):
\begin{equation}
\hat J_{ij}^{(c)}= n_{i,a}n_{j,a}+n_{i,b}n_{j,b}
 + a_i^\dagger b_i b_j^\dagger a_j
 + b_i^\dagger a_i a_j^\dagger b_j ~.
\end{equation}
Similar expressions can be obtained for $\hat J_{ij}^{\it (a)}$ and 
$\hat J_{ij}^{\it (b)}$ by replacing orbital fermions $a, b$ 
in this equation by $b, c$ and $c, a$ pairs, respectively. We notice, 
that in Eq.(2) and below fermions have only orbital quantum 
number ($a,b,c$ orbitons), since a spin component of the original 
fermions is represented in Eq.(1) by spin one-half operators, 
as usual. We shall focus now on the Hamiltonian (1), and discuss
later the effects of finite Hund coupling corrections (of the order of
$J_H/U$).

As evident from Eq.(1) (consider
$\left\langle{\bf S}_i{\bf S}_j\right\rangle =-1/4$), the classical 
N\'eel state is infinitely degenerate in orbital sector, reflecting 
that cubic symmetry is respected by the N\'eel state. 
This emphasizes a crucial importance of quantum effects, as has been
pointed out first in the context of $e_g$-systems \cite{fei97}. The bond 
directional geometry of $e_g$ orbitals offers a solution \cite{kha97} 
in that case: $e_g$ orbital frustration is resolved by order from disorder
mechanism selecting a particular (directional) orbital configuration,
which maximizes the energy gain from quantum spin fluctuations.
$t_{2g}$ orbitals are however not bond oriented (they are all 
planar), and not much spin fluctuation energy can be gained by any
pattern of static orbital orderings. The solution of the 
$t_{2g}$-problem proposed here is different: that is an
idea of $SU(4)$ spin-orbital excitations recently discussed in a 
context of one dimensional models \cite{li98,fri99,mil99}.

\begin{figure}
\epsfxsize=8cm
\centerline{\epsffile{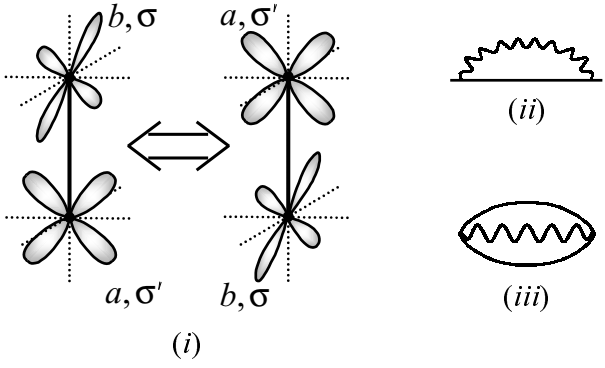}}
\vspace{5mm}
\caption{
({\it i}) Spin-orbital resonance between two (left and right) 
states of the {\it c}-bond pair. ({\it ii})Self-consistent 
Born approximation for the fermionic orbital self-energy.
({\it iii}) Diagram contributing to the bond order parameter $\chi$. 
Solid (wavy) lines represent orbitons (magnons).
}
\label{fig1}\end{figure}

Reflecting the geometry of $t_{2g}$ orbitals, every bond in Eq.(1)
is represented by two equivalent orbitals (such a symmetry is
absent for Ising-like $e_g$ orbitals). This is also evident
from the representation of Eq.(2) in terms of orbital pseudospin
${\bf \tau}_{ab}^i$ and density $n_{ab}^i = n_{ia} + n_{ib}$ 
operators acting on the $(a,b)$-doublet subspace: 
$\hat J_{ij}^{(c)}= 2\left( {\bf \tau}_{ab}^i {\bf \tau}_{ab}^j +
\frac{1}{4} n_{ab}^i n_{ab}^j \right).$
Suppose that $n_{ab}^i =1$; one then clearly observes an orbital 
$SU(2)$ symmetry of exchange ``integrals''. As we know from spin
$SU(2)$$\times$orbital $SU(2)$ models \cite{li98,fri99,mil99}, the exchange
energy is gained in that case due to the resonance between 
degenerate local configurations (spin-singlet $\times$ orbital triplet,
and spin-triplet $\times$ orbital singlet), and elementary excitations 
are a mixed spin-orbital $SU(4)$ modes. Fig.1{\it (i)} illustrates
the idea of correlated spin-orbital fluctuation. In the present problem,
spins and orbitals are different: spin sector is effectively half-filled
($n_{i,\uparrow}+n_{i,\downarrow}=1$) and must show a long-range order 
in 3D; on the other hand, the orbital occupation number 
$n_{ab}^i$ is not conserved but rather fluctuates around 
the average value 2/3. Because of the presence of third orbital, 
one has a less severe constraint 
$n_{i,a}+n_{i,b}+n_{i,c}=1$ in orbital sector. 
Even though a true $SU(4)$ excitation 
cannot develop here because of spin-orbital asymmetry, the analogy 
with 1D models suggests the way how to optimize 
the exchange energy: That is by the formation of a virtual
$SU(4)$ resonance which can be viewed in the present context as a 
local excitation composed of magnon and orbital fluctuation.
This resonance removes the orbital frustration dynamically,
and the disordered state of orbitals is precisely what is 
required to amplify this mechanism in all three directions 
of cubic lattice. 

Technically, the main ingredient of the theory is the three-particle
bond variable, 
$\hat \chi_\alpha^{(ij)} = \alpha_i^\dagger\alpha_j (s_i^\dagger + s_j ).$
Here, $\alpha$ denotes an appropriate fermionic orbiton, say $a$ or
$b$ orbitons for ${\it c}$ -bond pair ${\it (ij)}$, while 
$s^{\dagger}$ being
a magnon excitation about the N\'eel state. This is an analogy to  
$SU(4)$ flavor bond variable \cite{li98} in a spin ordered state. The bare
magnon dispersion is simply given by the Heisenberg interaction
with an average exchange constant
$J = \langle \hat J_{ij}^{(\gamma)} \rangle$, which 
will be calculated below. On the contrary, the orbital dynamics is 
exclusively due to the three-particle resonance, since spin-only
prefactor $\left\langle{\bf S}_i{\bf S}_j+1/4\right\rangle$ is 
almost zero ($\sim-0.05$) in 3D. We may present 
Eq.(1) as $H_{\rm SE} = H_{\rm sp} + H_{\rm int} + {\rm const.},$
where ${\it H_{sp}}$ describes bare magnons, and 
$H_{\rm int} = -\frac{1}{2}\sum_{\langle ij \rangle}\sum_{\alpha\beta}
\hat\chi_\alpha^{(ij)}\hat\chi_\beta^{(ji)}.$
The strategy is then to decouple ${\it H_{int}}$ in terms of bond variables.
The solution with uniform bond amplitude $\chi$ gives 
$H_{\rm int} \Rightarrow H_a + H_b + H_c + {\rm const.}$, where
\begin{equation}
H_a = -\chi\cdot\sum_{\langle ij\rangle_{bc}} 
\left\{ a_i^\dagger a_j \left( s_i^\dagger + s_j \right) + {\rm H.c.}\right\}
 ~. \end{equation}
Here the summation is restricted to $(b,c)$-planes.
Similar expressions hold for $b$ and $c$ orbitons. Since the 
filling factor is only 1/3 per orbital flavor, the orbital sector is far 
from nesting conditions, thus a mean-field uniform solution in  
the spirit of large N -theories \cite{aff88} is believed to be a 
reasonable starting point to describe a disordered state of orbitals.
Eq.(3) describes an effective hopping of orbitons accompanied by 
simultaneous magnon excitations. Thus, we have mapped the superexchange
in an undoped system to an effective ${\chi - J}$ model, which in a 
momentum space reads as
\begin{equation}
H_{\chi-J} = \sum_{\bf kp} \sum_{\alpha}M_{\bf kp}^{(\alpha)}
(\alpha_{{\bf k}+{\bf p}}^\dagger \alpha_{\bf k} s_{\bf p} +{\rm H.c.})
+\sum_{\bf p} \omega_{\bf p} s_{\bf p}^\dagger s_{\bf p} . 
\end{equation}
Here 
$M_{\bf kp}^{\alpha} = -4\chi (\gamma_{{\bf k}+{\bf p}}^{(\alpha)}u_{\bf p} + 
\gamma_{\bf k}^{(\alpha)}v_{\bf p})$, while form-factors are 
$\gamma_{\bf k}^{(a)} =(c_y+c_z)/2$, $\gamma_{\bf k}^{(b)} =(c_z+c_x)/2$, 
$\gamma_{\bf k}^{(c)} =(c_x+c_y)/2$, where $c_{\lambda}=cosk_{\lambda}$. 
Bare magnon dispersion $\omega_{\bf p} =3J(1-\gamma_{\bf p}^2)^{1/2}$ with
$\gamma_{\bf p} =(c_x+c_y+c_z)/3$, and $u_{\bf p}, v_{\bf p}$ are conventional Bogoliubov
transformation coefficients for magnons in 
the cubic lattice. It is noticed that the coupling constant in Eq.(4)
vanishes at $p=0$ limit (spin conservation is respected), and it is
the short-wavelength magnons which are important.
The model is similar to the  ${t - J}$ model for the doped N\'eel state 
of cuprates (see \cite{kan89,mar91} for comparison), yet the parameters
$\chi$ and $J$ have to be self-consistently determined. It is noticed,
that there are three branches of 2D fermions, 
and the dimensionality of magnon sector is three in the present case.

We calculate first the fermionic spectrum within a self-consistent Born 
approximation. Different from holes in the $t-J$ model, an orbiton 
creates a magnon on the site it arrives at, and 
eliminates a magnon while leaving the site (see Eq.(3)). Therefore 
orbiton motion contains a coherent component even in the Ising limit
for magnons. The latter approximation is used to simplify a momentum 
integration in the self-energy, Fig.1({\it ii}).
One then obtains $Re\Sigma_\omega^{(\alpha)}({\bf k}) =
f_{\omega}-\kappa_{\omega}(\gamma_{\bf k}^{(\alpha)})^2 $. 
Here $\kappa_\omega=(4\chi)^2 \int_0^\infty
d\xi \rho(\xi)/(\xi+3J-\omega)$, 
where $\rho(\xi)= \langle \rho_{\bf k}(\xi) \rangle_{\bf k}$ 
is fermionic density of states (DOS).  $f_\omega$
has similar structure: $f_\omega=(4\chi)^2 \int_{-\infty}^0
d\xi \tilde{\rho}(\xi)/(\omega+3J-\xi)$, 
where $\tilde{\rho}(\xi)= \langle \rho_{\bf k}
(\xi)(\gamma_{\bf k}^{(\alpha)})^2 \rangle_{\bf k}$.
One observes a two-sublattice 
structure of the orbiton dispersion (imposed by spin order) 
as expected from analogy with the $t-J$ model \cite{mar91}. We estimate
the fermionic (unrenormalized) mass as $m \simeq 1/\kappa_0 $, and density
of states as $\rho(0) \simeq 1/\pi\kappa_0$. Each orbiton forms its
own 2D fermi-surface (FS). Say, for $a$ orbitons the FS consists of two
almost circles around $(0,0)$ and $(\pi,\pi)$ points in a $(k_y,k_z)$
plane. Assuming constant DOS
within the interval of width $W_{orb}=1/\rho(0)$, we find
$\kappa_0 \simeq 4\chi /\sqrt{\pi}$, and
$W_{orb}\simeq 4\sqrt{\pi}\chi$. $W_{orb}$ is the energy scale of orbital 
fluctuations \cite{ref01}. Further, we estimate the bond amplitude     
$\chi = \langle a_i^\dagger a_j (s_i^\dagger + s_j) \rangle$. 
In the Ising limit
for intermediate magnons, Fig.1({\it iii}) gives $\chi \simeq 
(4/3\sqrt{\pi}) \langle (\gamma_{\bf k}^{(a)})^2 \rangle _{FS} $. 
For $\langle n^{(a)}\rangle =1/3$ we estimate 
$\langle (\gamma_{\bf k}^{(a)})^2 \rangle _{FS} \simeq 0.3$.
Spin stiffness in the present model is controlled by $J$. The mean-field
value 2/9 for $J$ follows from Eq.(2).This number is reduced to $J \simeq
0.16$ due to the corrections shown in Fig.2({\it i}). The physics 
behind this reduction is that $SU(4)$ resonance induces some 
ferromagnetic component in spin interactions.

Thus, we have fixed basic energy scales (in units of $4t^2/U$):
$W_{orb}\simeq1.6$ for orbital sector, and $3J\simeq0.48$ for magnon
bandwidth. The spin stiffness is relatively small because Pauli
principle is relevant only if electrons occupy the same orbital, and 
that probability is reduced at large orbital degeneracy. Main energy gain 
stems from magnon-orbital resonance: $E_{int}
=-6\chi^2 \simeq -0.31$. (Spin-only contribution is negligible: 
$E_{SW}\simeq -0.02$). This number should be compared with energy gain that 
one has with statically ordered planar orbitals: $\simeq -0.17$.

\begin{figure}
\epsfxsize=6cm
\centerline{\epsffile{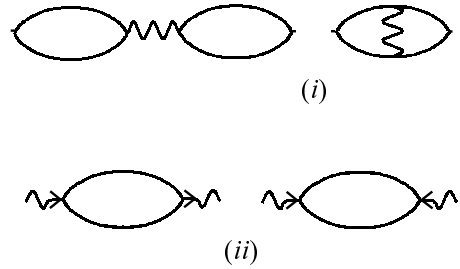}}
\vspace{5mm}
\caption{
({\it i}) Interaction corrections to the spin exchange constant $J$. 
({\it ii}) Magnon self-energies.
}
\label{fig2}\end{figure}

We discuss now physical consequences of the theory in 
context of experimental data on $LaTiO_3$ \cite{kei00}. 
Observed magnon dispersions of cubic symmetry are
very natural in above picture of orbitals fluctuating faster than
magnons. The energy scale $4t^2/U\simeq 100$ meV, required to fit
$J_{exp}=15.5\pm1$meV, is also reasonable in view of values 
$U\sim 4$eV \cite{miz96}, and $t\sim0.3$eV \cite{kat93}. This scale means that
orbital fluctuation energy in $LaTiO_3$ is $W_{orb}\sim160$meV.

Keimer et al. observed very small anisotropy gap, indicating that
effects of conventional relativistic spin-orbit coupling 
are strongly suppressed.
This is in fact an unavoidable consequence of our theory, since the 
ground state has no orbital degeneracy. We may calculate spectral 
density of orbital angular momentum fluctuations. It is given by 
the fermionic orbiton excitations (``orbiton Stoner continuum''). 
Spectral density of the local susceptibility of angular moments
 $l_i^z=i(a^{\dagger}b-b^{\dagger}a)_i$ vanishes in the 
static, $\omega=0$ limit:
$S(\omega)=2\rho^2(0)\omega$. This implies complete quenching of orbital 
moments at low energies. Making parallel with Jahn-Teller (JT) impurity 
physics \cite{ham65}, we may say that ``effective Ham reduction factor'' 
, $\zeta(\bar{\omega })\sim \left\{ \int_0^{\bar{\omega }}
S(\omega)d\omega \right\}^{1/2} \sim \bar{\omega }/W_{orb}$
is frequency dependent, and the angular moments disappear from low-energy
physics linearly in energy. Also, the ratio  
$\Lambda_{so}/W_{orb} \sim 10^{-1}$ ($\Lambda_{so} \sim 20$ meV\cite{kei00})
means that relativistic 
spin-orbit coupling induced corrections are small; we estimate 
the $g$-value shift $\Delta g/g\sim0.06$, consistent with \cite{kei00}. 

Despite three dimensionality of magnon spectra, the staggered moment
in $LaTiO_3$ is small, i.e. 0.45$\mu_B$ \cite{kei00} - an 
obvious problem for a 
spin-wave picture. The present theory resolves this difficulty:
The intensity of spin Bragg peak is partially taken away 
by a quantum magnon-orbiton resonance and redistributed over finite 
frequency region. In other words, fluctuating orbitals generate  
additional quantum spin fluctuations in the ground state.
We have calculated a spin 
moment reduction in a similar way as it was previously 
done in the $t-J$ model \cite{kha93}.  
Accounting for lowest order interaction corrections (Fig.2({\it ii}))
to the magnons, one finds:
\begin{eqnarray}
\delta S_{int}^z=\frac{3}{2}(4\chi)^2 &&\sum_{\bf kk'} n_{\bf k} (1-n_{\bf k'})
\Bigl\{ \frac{A_{\bf kk'}}{(\omega_{\bf p}+\xi_{\bf k'} -\xi_{\bf k})^2} \nonumber \\
&&- \frac{B_{\bf kk'}} {\omega_{\bf p} (\omega_{\bf p}+\xi_{\bf k'} -\xi_{\bf k})} \Bigr\},
\end{eqnarray}
where $\xi_{\bf k}$ is the $a$ orbiton dispersion, ${\bf k'} ={\bf k}+{\bf p}$, $A_{\bf kk'} =
\lambda_+^2 /(1+\gamma_{\bf p})+\lambda_+ \lambda_- /(1-\gamma_{\bf p}^2)^{1/2}$,
$B_{\bf kk'} =\lambda_+^2\gamma_{\bf p}/(1+\gamma_{\bf p})$, and 
$\lambda_\pm =\gamma_{\bf k}^{(a)} \pm\gamma_{\bf k'}^{(a)}$. By averaging first
the matrix elements $A,B$ in Eq.(5) over the Brillouin zone (they are 
rather regular functions), we evaluate $\delta S_{int}^z \simeq0.15$.
(Another estimation by using the Ising limit for magnons gives 0.13).
Adding also a conventional 3D spin-wave correction 
$\delta S_{SW}^z \simeq0.075$, one then obtains a staggered moment $0.55\mu_B$
in fair agreement with experiment.

The orbital liquid picture offers a simple explanation of puzzling Fano-type 
phonon anomalies observed by Raman scattering data in insulating titanates
(Reedyk et al.\cite{ree97}). These anomalies are most pronounced in $LaTiO_3$
, and we identify their origin as due to the coupling of phonons to orbital
excitations, noticing that the phonon position ($\sim$ 300 cm$^{-1}$) is right 
in the orbital Stoner continuum.

According to Goodenough-Kanamori rules, a local ferrotype orbital 
correlations are expected in a spin N\'eel state. A residual interactions 
(via magnons) between orbitons do indeed produce a triplet pairing
of orbitons within each fermionic branch. Treating the problem on 
BCS level, we find finite mean-field order parameter 
$\langle \alpha_i^\dagger\alpha_j^\dagger \rangle$ of p-wave 
symmetry (specifically,
of $p_x+ip_y$ -symmetry for $c$ orbitons, for instance), 
which opens an orbital gap. It is difficult
to calculate reliably the value of this gap (of the order of a few meV); we
predict however that a linear $\gamma$ -term in specific heat 
should be released
when a correlation gap in orbital spectrum is thermally washed out. 
We estimate $\gamma\sim40 mJ/mole K^2$. Specific heat measurements in $LaTiO_3$
up to $T\sim100K$ would be a crucial test for the theory.

From physical point of view, the proximity of  $LaTiO_3$ to Mott 
transition (charge gap is only 0.2 eV \cite{kat93}) is actually a key 
factor, which drives this compound to superexchange dominated
orbital liquid state. This state is stable, however, only if the 
effective orbiton bandwidth, $W_{orb} \sim 4t^2/U$, is large 
enough to suppress JT order. 
When one goes to another end compound, $YTiO_3$
(charge gap $\sim1$eV) the bandwidth narrows \cite{kat93}, thus reducing 
superexchange energy scale, and we expect static JT ordering of spatially 
more localized orbitals once the ratio $W_{orb}/E_{JT}$ becomes 
less than critical. With antiferrotype orbital ordering as observed 
in  $YTiO_3$ \cite{nak00} a classical
expectation value of Eq.(2) vanishes, and Hund's coupling $J_H$ term
induces a ferromagnetic state. The competition between the frustrated
superexchange on one hand, and JT plus Hund's couplings on the other hand 
is expected to result in the phase diagram shown in Fig.3. 

\begin{figure}
\epsfxsize=6.5cm
\centerline{\epsffile{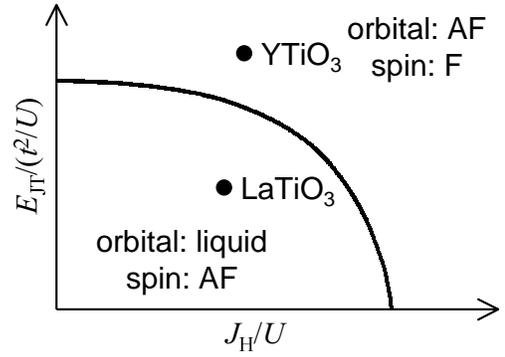}}
\vspace{5mm}
\caption{
Proposed phase diagram for $d^1(t_{2g})$ Mott insulator on cubic lattice.
At small Hund's and Jahn-Teller couplings the quantum N\'eel state with 
dynamically quenched orbital moments is stabilized. 
Quantum phase transition line separates this state from 
the ferromagnetic phase with static orbital/JT order.
}
\label{fig3}\end{figure}

To conclude, we have argued that due to large degeneracy and 
special geometry of orbitals $t_{2g}$-superexchange system 
most likely has an orbitally disordered ground state. The 
observed anomalies of $LaTiO_3$ find their natural explanations 
in a proposed orbital liquid picture. In metallic manganites, $e_g$
orbital disorder is enforced by the presence of mobile carriers 
\cite{ish97}.
Apparently, the case of insulating $LaTiO_3$ is a nice example of that 
a frustrated $t_{2g}$ orbitals can do the same job alone, without doping and 
well before the Mott transition to a metallic state is reached.

We would like to thank B. Keimer for stimulating discussions. 
Discussions with Y. Tokura, S. Ishihara, P. Prelov\v sek, and K. Tsutsui
are also acknowledged. This work was supported by Priority Areas Grants
from the Ministry of Education, Science, Culture and Sport of Japan,
CREST and NEDO.


\begin{references}
\bibitem{ref01}
More specifically, orbital dynamics contains both coherent and 
incoherent components. With $\chi, J$ given in the text, we have 
estimated the quasiparticle weight $\sim 0.7$, which is close to 
the ratio $J/\chi$.
\end{references}
\end{document}